\documentclass[aps,prd,showpacs,onecolumn,superscriptaddress,amsmath,amssymb]{revtex4}
\usepackage{epsfig}
\usepackage{bm}

\usepackage[english]{babel}

\usepackage{amsmath}
\usepackage{slashed}
\usepackage{braket}

\begin{document}
\title{Soft photons in semileptonic $B \to D$ decays}
\thanks{Presented by N. Ko\v snik at FLAVIAnet topical workshop: Low
  energy constraints on extensions of the Standard Model, Kazimierz,
  and Primosten09: Progress and challenges in flavour physics, Primo\v
  sten }

  \author{ Damir Be\'cirevi\'c} 
  \affiliation{Laboratoire de Physique Th\'eorique (B\^at. 210),
      Universit\'e Paris Sud,
      Centre d'Orsay, F-91405 Orsay-Cedex, France}
  
   \author{Nejc
      Ko\v{s}nik} \email[Electronic address:]{nejc.kosnik@ijs.si}
    \affiliation{J. Stefan Institute, Jamova 39, P. O. Box 3000, 1001
      Ljubljana, Slovenia}

\date{\today}

\begin{abstract}
  Determination of $V_{cb}$ in exclusive semileptonic decays is
  crucial consistency check against the $V_{cb}$ determined
  inclusively. Anticipated precision of $V_{cb}$ at the Super Flavor
  factory is $\sim 1\%$, with most of the theoretical error due to
  hadronic form factor uncertainties. However, at this level of
  precision treating electromagnetic effects becomes inevitable. In
  addition to virtual photon corrections there are also emissions of
  real photons which are soft enough to avoid detection. The
  bremsstrahlung part is completely universal and is accounted for in
  the experimental analyses. However, the so-called \emph{structure
    dependent} contribution, which probes the hadronic content of the
  process and is infrared finite, has been neglected so far. To this
  end, we estimated fraction of radiative events which are identified
  as semileptonic by experiment.
\end{abstract}
\pacs{13.30.Ce,13.40.Ks}

\maketitle

\section{Introduction}
Many efforts have been devoted to experimentally check the validity of
Kobayashi-Maskawa~(KM) mechanism which predicts that all quark flavor
observables agree with the unitary CKM matrix and single CP violating
phase. The KM mechanism states that either measuring sides
or angles of the unitarity triangle, the apex $(\bar\rho,\bar\eta)$
comes out unique. Value of $V_{cb}$ determines lengths of sides
adjacent to the apex, among them also the side opposite to angle
$\beta$ which is precisely measured. Current average of inclusive and
exclusive determinations is~\cite{pdg}
\begin{equation}
  |V_{cb}| = (41.2\pm 1.1)\times 10^{-3},
\end{equation}
where $|V_{cb}|_{excl} = (38.6\pm 1.3)\times 10^{-3}$ is significantly
lower than $|V_{cb}|_{incl} = (41.6\pm 0.6)\times 10^{-3}$. Common
lore is that most of theoretical error of the exclusive method stems
from the $B\to D$ form factors uncertainties and detection
efficiencies. Although inclusive analyses are under better control
theoretically and consequently result in more precise result,
exclusive method provides a crucial cross-check, since errors are
believed to be largely independent for both methods. Future
expectation for the exclusive precision is about $1\%$ which could be
obtained at Super Flavor factory~\cite{Browder:2008em}.

\section{Determination of $V_{cb}$ in $B \to D \ell \nu$}
\subsection{Theory input}
Differential rate of exclusive decay to pseudoscalar $D$ is
\begin{equation}
  \frac{d\Gamma}{dw} (B \to D \ell \nu) = \frac{G_f^2 |V_{cb}|^2}{48 \pi^3} (m_B+m_D)^2 m_D^3 (w^2-1)^{3/2} \mathcal G(w)^2,
\end{equation}
where $w= v\cdot v'$ is the scalar product of meson velocities. Heavy
quark symmetry normalizes the form factor $\mathcal G(w)$ at the
maximum recoil point ($w=1$), where final state $D$ meson is at rest
in the $B$ rest frame. Perturbative $\alpha_s$, $\alpha_{em}$, and
nonperturbative $(\Lambda_{QCD}/m_b)^n$ symmetry breaking corrections
were also computed and are under control at the maximum recoil
point. However, further theoretical insight is required to isolate the
value of $V_{cb}$. Allowed phase space shrinks as $w$ approaches $1$
and there are very few events recorded in this region. So to infer the
experimental value of $V_{cb} \times \mathcal G(1)$ one has to rely on
a particular shape of the form factor to guide the extrapolation down
to $w=1$. In experimental literature it has become customary to use
so-called CLN shapes of the form factors~\cite{Caprini:1997mu} which
rely on analyticity and unitarity. Measured differential decay rate is
then fitted with $V_{cb} \times \mathcal G(1)$ and slope $\rho^2$ of
the form factor at $w=1$. In the end theoretical prediction of
$\mathcal G(1)$ is used to determine $V_{cb}$.

\subsection{Experimental method}
Measuring semileptonic $B \to D \ell \nu$ in $e^+ e^-$ collider
operating at $\Upsilon(4s)$ resonance one can focus only on events
where the tag side momentum is completely reconstructed and ensure
that missing invariant mass is peaking at zero, as anticipated for a
single neutrino in final state. Kinematical constraints are applied
with some tolerance (invariant mass of the tag side is $5.27 -
5.29\,\mathrm{GeV}$ for decay of $B^-$, c.f.~\cite{Aubert:2009ac})
which allows the soft photon events to be included among the
semileptonic events.

In this study we set out to study radiative corrections of
semileptonic decay $B \to D \ell \nu$ and in particular what is the
ratio of structure dependent~(SD) radiative to semileptonic events
numbers for given photon energy cut of the experiment. We will keep
only the lowest pole contributions in our treatment as they turn out
to contribute dominantly due to kinematics. Similar studies were
carried out for $K$ meson semileptonic decays using chiral
perturbation theory, and as it had turned out SD part was negligible
for a typical experimental setup~\cite{Gasser:2004ds}. On the
contrary, SD amplitude of $B \to \mu \nu \gamma$ can lead to $\sim
20\%$ background in a typical experiment measuring $Br(B \to \mu
\nu)$~\cite{Becirevic:2009aq}.

\section{Infrared electromagnetic corrections}
Electromagnetic effects render all experimentally measured widths a
sum of rate of specific process plus rates of radiative events with
final
state photons which cannot be resolved by the
experiment. Such inclusive and infrared~(IR) safe quantity is schematically
\begin{equation}
  d\Gamma_{exp}(i \to f) =  d\Gamma(i \to f) + d\Gamma(i \to f \gamma)_{E_\gamma < E_\mathrm{cut}} + d\Gamma(i \to f \gamma\gamma)_{E_{\gamma i}<E_\mathrm{cut},\,\sum E_{\gamma i}<E'_\mathrm{cut}} + \cdots.
\end{equation}
The above inclusive width solves the IR problem of electrodynamics by
cancelling soft divergences due to virtual photon corrections against
real emission. The amplitude of the so-called inner
bremsstrahlung~(IB) diverges as the photon energy approaches zero and
residue of the pole is fixed by the charge of the external leg where
the photon is emitted from. In the IR limit photons can only resolve
total charge of the emitting particle. Accordingly, Low's theorem
states that leading two terms in momentum expansion of the radiative
decay width are fixed from the nonradiative decay
width~\cite{Low:1958sn}. These IR divergences are compensated by the
corresponding virtual corrections at the same order of $\alpha_{EM}$
at the level of decay width.

However, there are also subleading, IR finite, terms in the
$d\Gamma_{exp}$ which are usually neglected in experimental analyses.
These structure dependent photon emissions can resolve structure
of charged particles. Consequently, prediction of SD terms require
knowledge of additional form factors.

\subsection{Amplitude}
We adopt notation established in~\cite{Gasser:2004ds} for semileptonic
$K$ decays. Amplitude of $B^- \to D^0 \ell \nu \gamma$ is
\begin{equation}  
\label{amp}
  \mathcal{A}_\mu = \frac{e G_F V_{cb}}{\sqrt{2}} \, \bar
  u(p_l) \left(-\frac{F_\nu(t)}{2 p_\ell \cdot q} \gamma_\mu
    (\slashed{p}_\ell+\slashed{q}+m_\ell) +
    V_{\mu\nu}-A_{\mu\nu} \right) \gamma^\nu (1-\gamma_5) v(k)
\end{equation}
which is in the end contracted with the photon polarization. Here $q$,
$k$, and $p_\ell$ are the respective photon, neutrino, and lepton momenta.
First term in brackets is proportional to
\begin{equation}
  F_\nu(t) \equiv i \Braket{D(p') | H_\nu|B(p)},\qquad t \equiv (p-p')^2
\end{equation}
and represents the photon emission from the lepton leg, whereas
$V_{\mu\nu}$ and $A_{\mu\nu}$ are hadronic vector and axial form
factors of $B \to D\gamma$ transition, namely when photon is emitted
from hadronic line
\begin{equation}
  V_{\mu\nu}-A_{\mu\nu} \equiv \int d^4y\,e^{i q \cdot y}
  \Braket{D(p') |T\left[J_\mu(y) H_\nu(0) \right]|B(p)}, \quad H^\nu \equiv \bar c \gamma^\nu (1-\gamma_5) b.
\end{equation}
Here $J_\mu$ is the electromagnetic current. These form factors
obey electromagnetic Ward identities
\begin{subequations}
\begin{align}
   q^\mu V_{\mu\nu} &= F_\nu(t),\\
 q^\mu A_{\mu\nu} &= 0,
\end{align}
\end{subequations}
which ensure total amplitude is gauge invariant. Intermediate
1-particle resonances give rise to poles due to excited beauty and
charm states. The soft photon part of phase space should be well
approximated by lowest pole contributions of $B$, $B^*$ and $D^*$.
The $B$-pole satisfies the inhomogeneous Ward identity above and
we single it out of $V_{\mu\nu}$
\begin{align*}
  V_{\mu\nu}^{IB} &= \frac{p_{\mu}}{p\cdot q} F_\nu (t)\\
  V_{\mu\nu}^{SD} &= V_{\mu\nu}-V_{\mu\nu}^{IB}, \quad q^\mu V_{\mu\nu}^{SD} = 0.
\end{align*}
Lorentz covariance and Ward identities allow the form factors to be
split down into eight scalar functions $V_{1\ldots 4}, A_{1\ldots 4},$
\begin{subequations}
\begin{align}
  V^{SD}_{\mu\nu} =& V_1 \left(p'_\mu q_\nu- p'\cdot q\,
    g_{\mu\nu}\right)
  + V_2 \left(p_\mu q_\nu  - p\cdot q\, g_{\mu\nu}\right)\nonumber\\
  &+(p\cdot q \,p'_\mu - p'\cdot q\,p_\mu)
  \left(V_3\, p_\nu +V_4\, p'_\nu\right),\\
  A_{\mu\nu} =& A_1 \epsilon_{\mu\nu\alpha\beta}\,
  p^\alpha q^\beta +A_2 \epsilon_{\mu\nu\alpha\beta}\, p'^\alpha
  q^\beta +\left(A_3 p_\nu+ A_4
    p'_\nu\right)\,\epsilon_{\mu\alpha\beta\gamma}\, p^\alpha q^\beta
  p'^\gamma.
\end{align}
\end{subequations}
We saturate the SD part of the amplitude with $D^*$ and $B^*$
resonances, which contribute to $V_{\mu\nu} - A_{\mu\nu}$ as
\begin{subequations}
\label{poles}
\begin{align}
  &\frac{i \Braket{D | J_\mu | D^*(p'+q)} \Braket{D^*(p'+q) | V_\nu-A_\nu | B}}{(p'+q)^2-m_{D^*}^2},\\
  &\frac{i \Braket{D | V_\nu-A_\nu | B^*(p-q)}
    \Braket{B^*(p-q) | J_\mu | B}}{(p
    -q)^2-m_{B^*}^2}.
\end{align}
\end{subequations}
The $B^*$ pole is not far in unphysical region and its contribution
gets enhanced by factor $1/(m_B^2-m_{B^*}^2)$ in the limit $E_\gamma
\to 0$. The $D^*$ pole, on the other hand, can be on-shell and we
model its contribution by Breit-Wigner shape
\begin{equation}
  \frac{i \Braket{D | J_\mu | D^*} \Braket{D^* | V_\nu-A_\nu | B}}{(p'+q)^2-m_{D^*}^2+i m_{D^*} \Gamma_{D^*}}.
\end{equation}
The above on-shell $D^*$ contribution is expected to dominate the
radiative decay in question. Form factors $V_{\mu\nu},A_{\mu\nu}$
contain nonperturbative matrix elements, as evident
from~(\ref{poles}), for which we take quenched lattice results of $B
\to D^{*}$ form factors~\cite{deDivitiis:2008df,deDivitiis:2007uk}.
Value of $g_{D^*D\gamma}$ was computed on the lattice with dynamical
light quarks~\cite{Becirevic:2009xp}.

\begin{figure}[!h]
 \begin{center} 
\epsfig{file=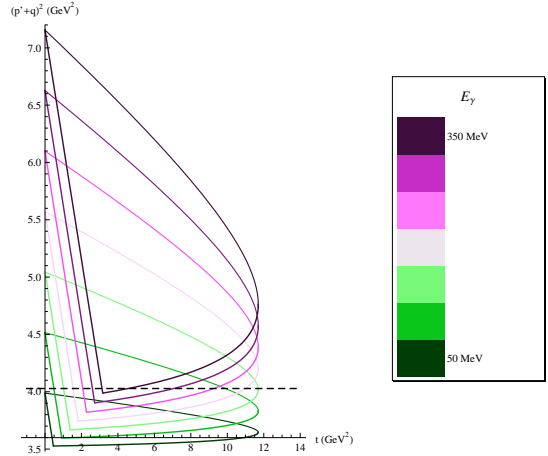,height=6cm}
\end{center}
\caption{Slices of phase space in the $D^*$ invariant mass versus
  momentum transfer $t$ for different photon energies. Horizontal line
  represents on-shell $D^*$, which is reachable only in the range
  $50\,\mathrm{MeV} < E_\gamma < 350\,\mathrm{MeV}$. }
\label{kinematics}
\end{figure}
The intermediate $D^*$ is kinematically allowed to be on-shell only
for photon energies in the range of $\sim [50,350]\,\mathrm{MeV}$~(see
Fig.~\ref{kinematics}). This resonant enhancement of the soft photon
kinematical region originates from relatively small mass splitting
between $D^*$ and $D$. Next higher excited charm state lies already
above $2.4\,\mathrm{GeV}$ and would result in more energetic
photons due to larger mass splitting. Thus we expect that higher
resonances would mostly produce photons in the experimentally
accessible region.

\begin{figure}[!h]
\begin{center}
\begin{tabular}{cc}
\epsfig{file=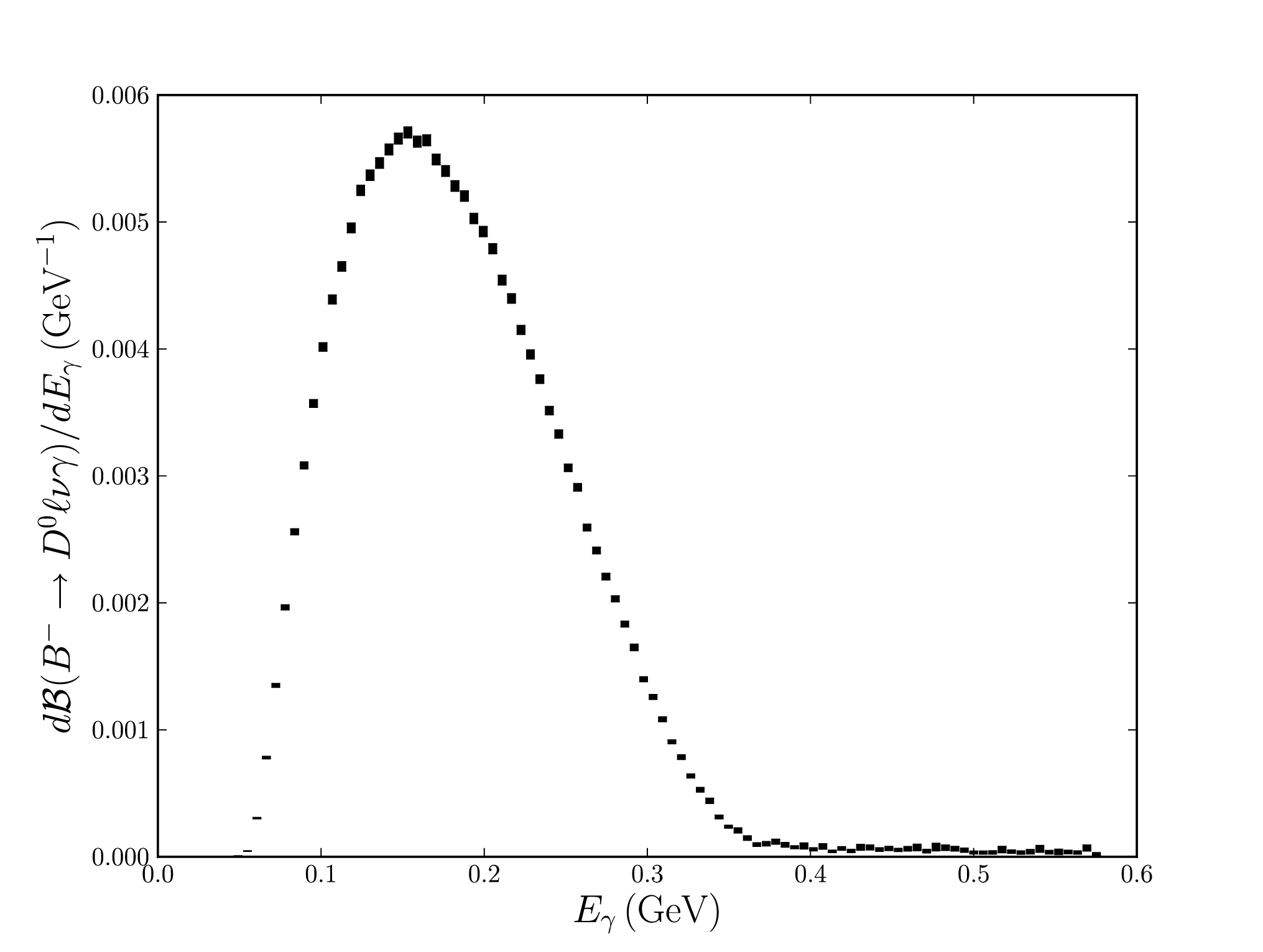,height=6.5cm} & \epsfig{file=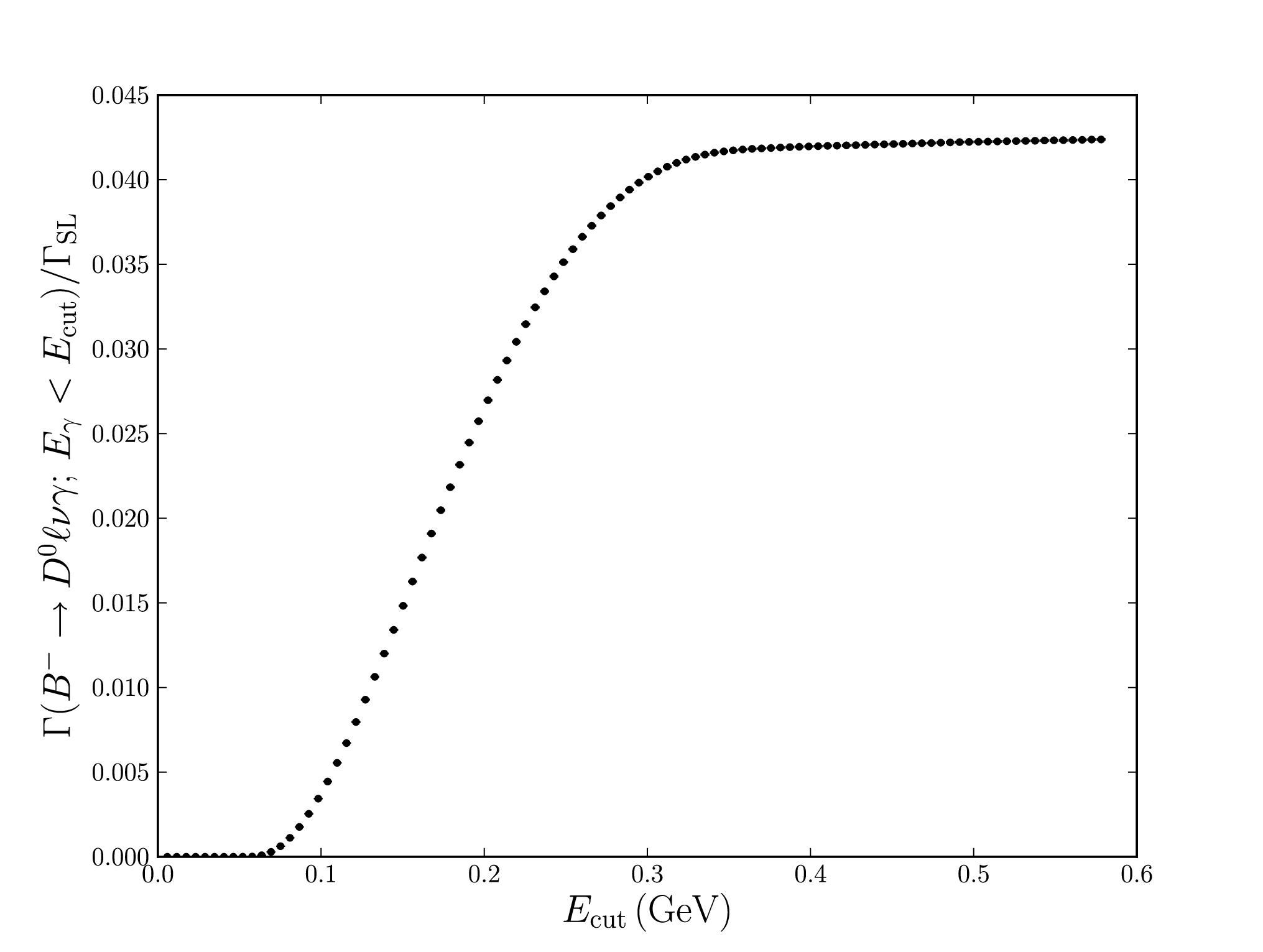,height=6.5cm}
\end{tabular}
\end{center}
\caption{Left: resonant $D^*$ spectrum of $B^- \to D^0 \mu \nu
  \gamma$. Right: fraction of misidentified radiative events plotted
  against the experimental resolution of the photon energy $E_{cut}$.}
\label{muspectrum}
\end{figure}

\begin{figure}[!h]
\begin{center}
\begin{tabular}{cc}
  \epsfig{file=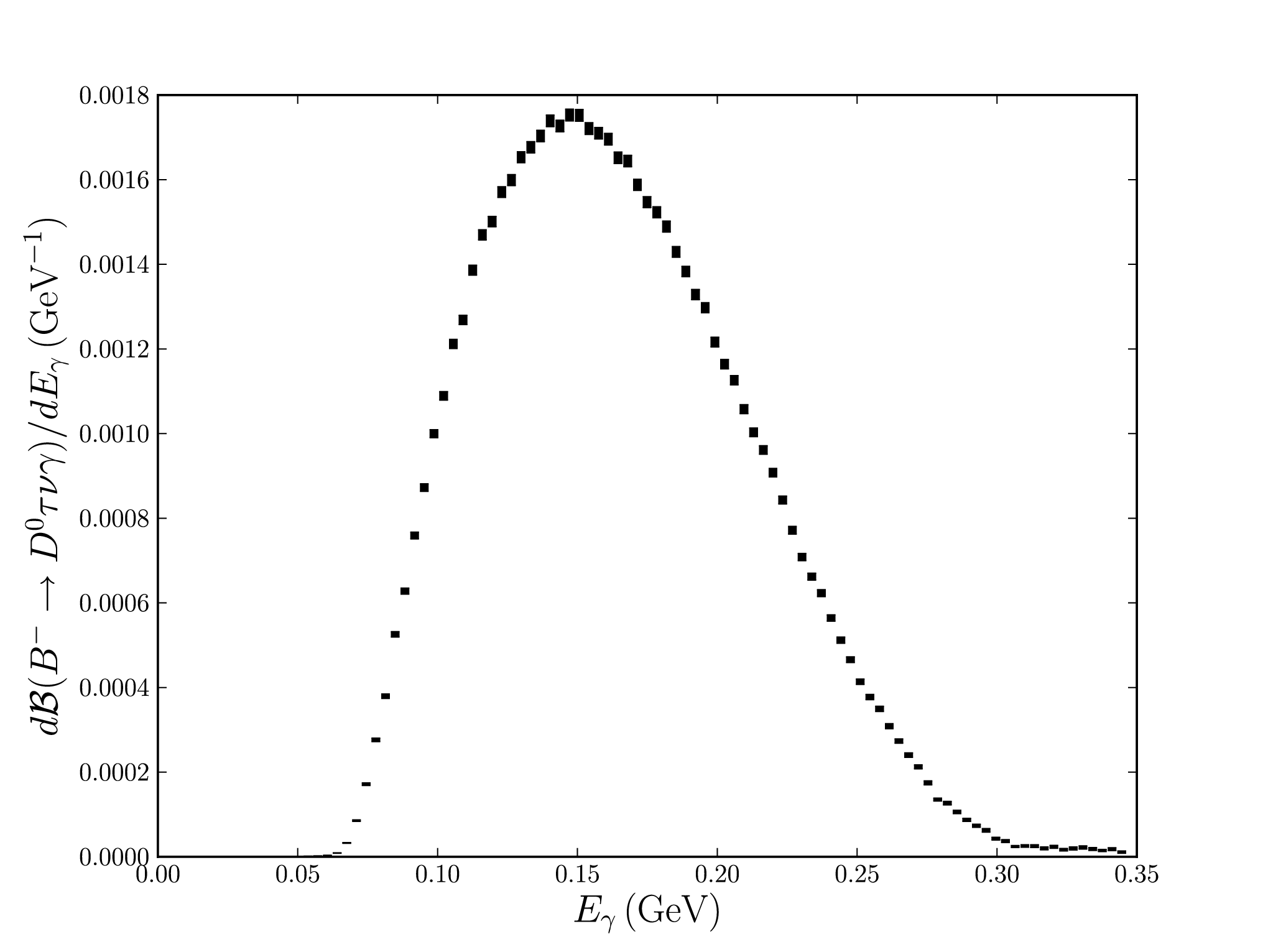,height=6.5cm} & \epsfig{file=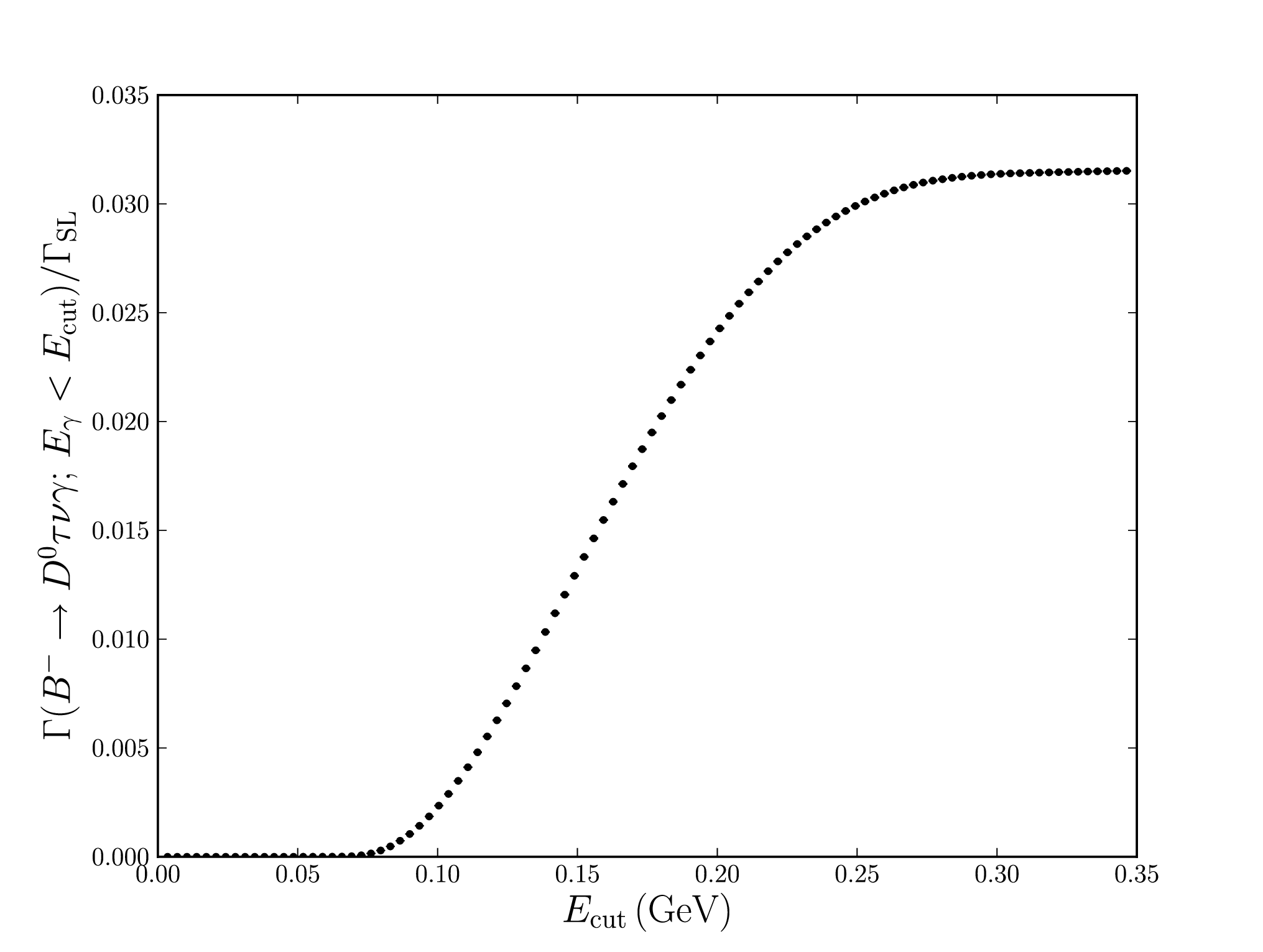,height=6.5cm}
\end{tabular}
\end{center}
\caption{Left: resonant $D^*$ spectrum of $B^- \to D^0 \tau \nu
  \gamma$. Right: fraction of misidentified radiative events plotted
  against the experimental resolution of the photon energy $E_{cut}$.}
\label{tauspectrum}
\end{figure}

The contribution of $D^*$ is clearly seen in the $E_\gamma$ spectra
of $\mu$ and $\tau$ channels,
Figs.~\ref{muspectrum},\ref{tauspectrum}, where roughly half of the
width lies in the $E_\gamma < 200\,\mathrm{MeV}$ region.

The importance of improving experimental resolution of soft photons
detection or including them among backgrounds is clearly seen in
Figs.~\ref{muspectrum}, \ref{tauspectrum}. A photon energy cut of
$300\,\mathrm{MeV}$ in $B^- \to D^0 \mu \nu$ results in $\sim 4\%$ of
the recorded events to be fake. This would in turn imply a $2\%$ fake
enhancement of $|V_{cb}|$.

\section*{Acknowledgements} 
N.K. would like to thank LPT Orsay for hospitality and financial
support during his stay in Paris, where part of this work was
done. This work is partially funded by Slovenian research agency.



\end{document}